\PassOptionsToPackage{hyphens}{url} %

\documentclass[letterpaper,twocolumn,10pt]{article}
\usepackage{usenix-2020-09}

\usepackage[normalem]{ulem}
\usepackage[inline]{enumitem} %
\usepackage{microtype} %
\usepackage{booktabs} %
\usepackage{amssymb} %
\usepackage{pifont} %
\usepackage{algorithm} %
\usepackage{algpseudocode}
\usepackage{dutchcal}

\usepackage{footnote} %
\makesavenoteenv{table}
\makesavenoteenv{tabular}
\makesavenoteenv{tabu}

\usepackage{tikz}
\usepackage{xcolor}
\definecolor{node}{HTML}{808080}
\newcommand*\boxed[1]{\tikz[baseline=(char.base)]{
        \node[shape=rectangle,fill=node,text=white,scale=0.9,inner sep=2,outer sep=2,minimum width=10pt,font=\fontfamily{phv}\selectfont] (char) {#1}}}

\usepackage{tcolorbox}
\newtcolorbox{observationbox}[1][]
{
    left=0pt,right=0pt,top=0pt,bottom=0pt,
    #1
}

\microtypecontext{spacing=nonfrench}

\newcommand{\footurl}[1]{\footnote{\url{#1}}}
\newcommand{\yes}[0]{\ding{51}}
\newcommand{\no}[0]{\ding{55}}
\newcommand{\app}[1]{\mbox{\hyperref[app:#1]{App-#1}}}
\newcommand{\definition}[1]{\hyperref[def:#1]{Definition~#1}}

\begin{document}

\date{}  %

\title{\Large Let's Trace It: \bf Fine-Grained Serverless Benchmarking using \\ Synchronous and Asynchronous Orchestrated Applications}

\author{
{\rm Joel~Scheuner}\\
Chalmers | University of Gothenburg\\
Sweden\\
scheuner@chalmers.se
\and
{\rm Simon~Eismann}\\
University of Würzburg\\
Germany\\
eismann@uni-wuerzburg.de
\and
{\rm Sacheen~Talluri}\\
Vrije Universiteit Amsterdam\\
The Netherlands\\
S.Talluri@atlarge-research.com
\and
{\rm Erwin~van~Eyk}\\
Vrije Universiteit Amsterdam\\
The Netherlands\\
E.vanEyk@atlarge-research.com
\and
{\rm Cristina L. Abad}\\
Escuela Superior Politecnica del Litoral\\
Ecuador\\
cabadr@espol.edu.ec
\and
{\rm Philipp Leitner}\\
Chalmers | University of Gothenburg\\
Sweden\\
philipp.leitner@chalmers.se
\and
{\rm Alexandru~Iosup}\\
Vrije Universiteit Amsterdam\\
The Netherlands\\
A.Iosup@atlarge-research.com
} %

\maketitle

\begin{abstract}
Making serverless computing widely applicable requires detailed performance understanding. %
Although contemporary benchmarking approaches exist,
they report only coarse results, do not apply distributed tracing, do not consider asynchronous applications, and provide limited capabilities for (root cause) analysis.
Addressing this gap, we design and implement ServiBench, a serverless
benchmarking suite. 
ServiBench 
(i) leverages synchronous and asynchronous serverless applications representative of production usage, 
(ii) extrapolates cloud-provider data to generate realistic workloads, 
(iii) conducts comprehensive, end-to-end experiments to capture application-level performance, 
(iv) analyzes results using a novel approach based on (distributed) serverless tracing, and
(v) supports comprehensively serverless performance analysis.
With ServiBench, we conduct comprehensive experiments on AWS, covering five common performance factors: median latency, cold starts, tail latency, scalability, and dynamic workloads.
We find that the median end-to-end latency of
serverless applications is often dominated not by function computation but 
by external service calls, orchestration, or trigger-based coordination. 
We release collected experimental data under FAIR principles and ServiBench as a tested, extensible open-source tool.
\end{abstract}

\section{Introduction}\label{introduction}

Establishing computing infrastructure that is easy to manage yet performs well for all applications is a longstanding goal of the computer systems community from the 1950s~\cite{book/Yost17}. Emerging in the late 2010s from the integration of multiple technological breakthroughs~\cite{DBLP:journals/internet/EykTTVUI18}, 
\textit{serverless computing~}\cite{castro:19a,eyk:17,DBLP:journals/cacm/Schleier-SmithS21} aims to abstract away operational concerns (e.g., autoscaling) from the developer by providing fully managed cloud platforms through self-serving application programming interfaces (APIs).
Developers can leverage a rich \textit{ecosystem of external services} (e.g., message queues, databases, image recognition), which are glued together by a Function-as-a-Service~(FaaS) platform, such as AWS Lambda.
For the current generation of serverless platforms, ease of management comes with important performance trade-offs and issues, including high tail-latency and performance variability~\cite{pelle:19,gan:19}, and delays introduced by asynchronous use of external services~\cite{pelle:19,quinn:21}. 
Thus, understanding and comparing the performance of serverless platforms is essential.
Although extensive prior work exists in empirical performance evaluation~\cite{figiela:18,manner:18,lee:18} and analysis~\cite{wang:18,kuhlenkamp:20}, synthetic~\cite{gan:19,ferdman:12} and micro-benchmarking~\cite{scheuner:20-jss}, as well as server-side benchmarking~\cite{shahrad:19,ustiugov:21}, there currently exists no serverless benchmark
that provides detailed (white-box) analysis at application level,
covering production applications and invocation patterns.
Addressing this gap, 
in this work we design, implement, and use \textit{ServiBench~(sb)}, an application-level, serverless benchmarking suite based on distributed tracing.
We posit in this paper that \textbf{a serverless benchmark that provides performance information at application level is necessary}. Our argument is two-fold. %
First, a variety of production-ready serverless applications already exist~\cite{leitner:19,castro:19a,eismann:21-tse}. 
These applications have distinctive performance profiles but alternative approaches insufficiently cover this diversity (e.g., in external services).
Second, application-level benchmarks have proven useful in related fields (e.g., 
DeathStarBench~\cite{gan:19} for container-based microservices, and CloudSuite~\cite{ferdman:12} for scale-out workloads). They help identify architectural bottlenecks, guide application design decisions, and inspire better programming models.

The challenge of application-level serverless benchmarking is manifold and complex. %
First, there is a \textbf{need to design the benchmarks, and to validate the tools that realize the benchmark in practice (challenge C1)}.
The majority of existing work attempts to reverse-engineer commercial serverless systems~\cite{wang:18}, by characterizing performance aspects such as allocated CPU power by memory size~\cite{figiela:18}, cold start overhead~\cite{manner:18}, scaling policies~\cite{kuhlenkamp:20}, or I/O speed~\cite{lee:18}.
Such studies can guide cloud users in selecting appropriate services and configurations, but their empirical results are prone to become obsolete quickly.
Server-side experimentation~\cite{shahrad:19,ustiugov:21} allows to control the entire serverless stack to obtain detailed profiling data, but lacks tight integration with external services, which prevents realistic system-level testing of serverless applications~\cite{lenarduzzi:20}. Finally, micro-benchmarks  (e.g., of CPU performance) are not representative of real applications.
Summarizing, no benchmark currently:
\begin{enumerate*} [label=\upshape(\roman*\upshape)]
     \item supports a variety of architectural patterns that appear commonly in serverless applications,
     \item includes representative, production-grade applications and invocation patterns,
     \item provides end-to-end performance data and white-box analysis capabilities, and 
     \item enables reproducible real-world experiments.
\end{enumerate*}

Second, \textbf{conducting white-box analysis requires collecting end-to-end performance data, and extracting fine-grained latency information (C2).} 
Distributed tracing~\cite{mace:17,sambasivan:14} has been popularized at Google~\cite{sigelman:10} and Facebook~\cite{veeraraghavan:18}, but requires a level of cooperation with the platform that is not available for serverless applications.
Various approaches exist for this broad class of problems that assume ordered events and accurate timing, and such approaches can already be useful for \textit{synchronous} microservice architectures~\cite{qiu:20}. In contrast, many serverless applications use multiple functions and external services, and invocations can occur \textit{asynchronously}. No benchmark in the field currently addresses this.

Third, we identify the \textbf{need to share tools and data for serverless benchmarking, FAIRly (C3).}
Releasing software, data, and results in packages that make them FAIR~(``findable, accessible, interoperable, and reusable''~\cite{data:FAIR16}) is essential in science and engineering.

We address these challenges with a four-fold contribution:
\begin{enumerate}
    \item \textbf{Application benchmark suite (Section~\ref{sec:benchmark_suite}):}
    Addressing C1, we present ServiBench, a comprehensive benchmark suite. In it, 10~ realistic open-source applications cover different forms of orchestration, synchronous and asynchronous triggers, and real-world characteristics such as programming language, size, and external service usage.
    ServiBench orchestrates reproducible deployments, automates realistic load generation, collects distributed traces, and provides detailed white-box analysis.

    \item \textbf{Latency breakdown analysis (Section~\ref{sec:trace_analysis}):}
    Addressing C2, 
    we design novel algorithms and heuristics for detailed latency breakdown analysis of distributed serverless traces. The key capability over prior work is that our approach works in a serverless context, across asynchronous call boundaries and external services.

    \item \textbf{Empirical performance study (Section~\ref{sec:results}):}
    Addressing the validation aspects of C1, and the overall challenge of understanding the performance of serverless applications, 
    we conduct a comprehensive white-box analysis of serverless application performance %
    in the AWS environment~(see also Section~\ref{sec:related_work}). Our results cover, e.g., cold starts, tail latency, and the impact of application type and invocation patterns on performance.

    \item { \textbf{FAIR release} of the ServiBench software, data, and results~(Appendix~\ref{app:replication})}:
    Addressing C3, we release ServiBench on Github, and the configurations and full data ($\pm50$\,GB) %
    on Zenodo (links anonymized).
\end{enumerate}

\section{System Model for Serverless Applications}\label{sec:system_model}

\begin{figure}[t]
    \centering
    \includegraphics[width=0.45\textwidth]{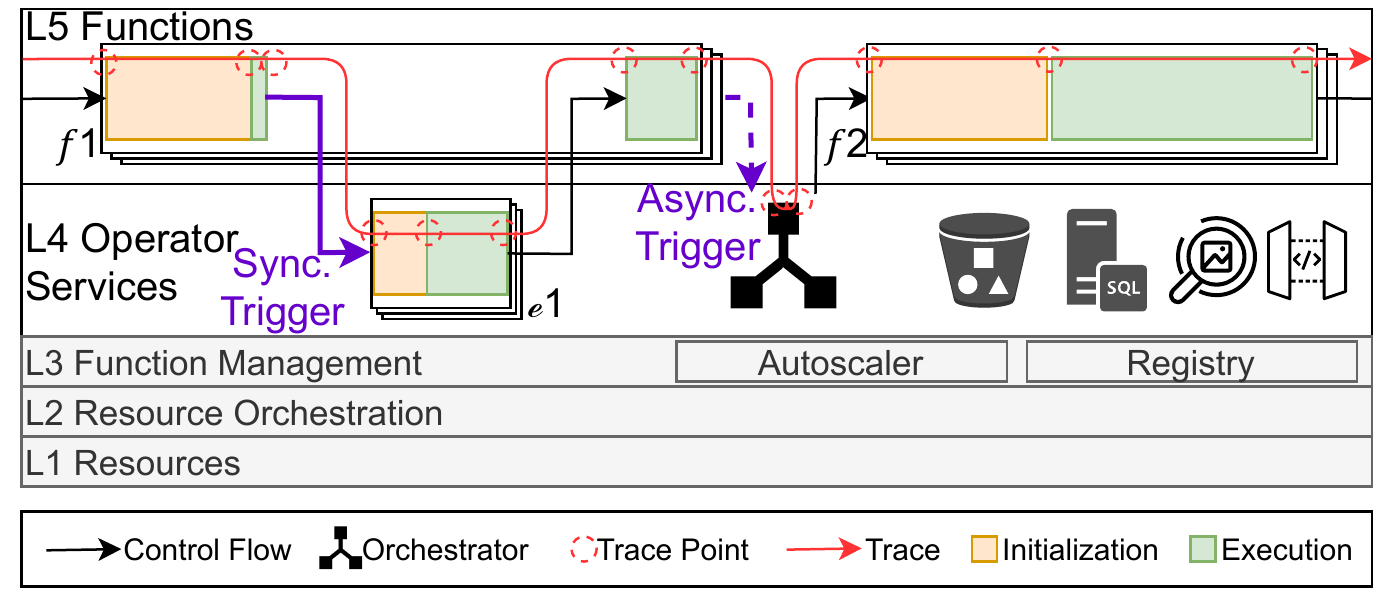}
    \vspace{-6pt}\caption{System model of a serverless application composed of multiple functions: $\mathcal{f}1$ and $\mathcal{f}2$ are user-defined functions, and $\mathcal{e}1$ is an external service.}
    \label{fig:system_model}
\end{figure}

Our work assumes a system model, introduced by the SPEC Research Group, that represents the operation of tens of existing serverless platforms~\cite{DBLP:journals/internet/EykIGEBVTSHA19}. In this model, the serverless stack provides resources~(label~\boxed{L1} in Figure~\ref{fig:system_model}), whose use is orchestrated~(\boxed{L2}), enabling complex function management~(\boxed{L3}) such as auto-scaling and image-registry. These elements have been covered extensively by the community~\cite{shahrad:20,DBLP:conf/fast/AnwarMTLRCZNWLH18}. %

The system model further includes two layers that service directly the application and are the focus of this work. %
Each applications is composed of a single or, more commonly, multiple user-defined functions~(\boxed{L5}), which can trigger asynchronously external (operator-provided) services~(\boxed{L4}).
Figure~\ref{fig:system_model} depicts a constructed example of a serverless application. 
There are two user-defined functions ($\mathcal{f}1$ and $\mathcal{f}2$).
During its execution, $\mathcal{f}1$ triggers synchronously an operator-provided service, $\mathcal{e}1$. At the end of its execution, $\mathcal{f}1$ triggers asynchronously an operator-provided orchestrator service.

\textit{Production-level serverless applications} fit this model well. The functions can communicate with external services~($\mathcal{e}1$), and with each other, directly via synchronous or asynchronous triggers. They can also communicate via an orchestrator, which decides on the control flow based on user-provided instructions.
They can thus construct complex execution paths, using any of the operator-provided services, e.g., object stores, databases, and ML services. 
A \textit{production workload} can  incorporate multiple invocation patterns, leading to complex performance behavior that incorporates not only intra-application latency, but also cross-application delays.

We further assume there exists a \textit{monitoring service}, which can provide detailed performance information and in particular trace the upper layers in the serverless stack (i.e., \boxed{L4} and \boxed{L5}). We do \textit{not} assume that the other layers are observable by the application; e.g., there is no server-side tracing. These assumptions match the operation of common serverless platforms, such as AWS. 
Although we assume the presence of \textit{distributed tracing}, we do \textit{not} assume its results are consistent and ordered across multiple components in the system or that the application triggers can only occur synchronously. This is consistent with how popular serverless platforms, such as AWS and Microsoft Azure, operate in practice. 

\section{Design of ServiBench}\label{sec:benchmark_suite}

In this section, we design ServiBench, an application-level benchmarking suite.
We design around a set of 5 design principles, creating a novel architecture and benchmarking flow. The suite includes 10 representative, open-source applications. ServiBench includes a process to generate representative workloads from existing invocation logs provided by cloud operators such as Microsoft Azure.

In the system model introduced in Section~\ref{sec:system_model}, identifying the latency contribution of each service to the total response time of the application is challenging. 
In particular, an application can take multiple paths of execution and trigger asynchronously different methods depending on runtime conditions, which breaks straightforward distributed tracing. We address this concern through a detailed design, in Section~\ref{sec:trace_analysis}.

\subsection{Design Principles} %
Based on guidelines on benchmarking best-practices~\cite{kistowski:15,hasselbring:21} and inspired by the microservice benchmark suite DeathStarBench~\cite{gan:19}, we formulate the following design principles:
\begin{enumerate}
    \item \textbf{Representativeness:}
    A representative and relevant benchmark suite closely matches the characteristics of real-world applications.
    We select applications from industrial workshops and academic studies based on the most common serverless
    application types~\cite{eismann:21-tse,castro:19a},
    programming languages~\cite{leitner:19,eismann:21-tse},
    application sizes~\cite{shahrad:20,leitner:19,eismann:21-tse},
    and external services~\cite{shahrad:20,leitner:19,eismann:21-tse}.
    \item \textbf{End-to-end operation:}
    An application-level serverless benchmark should implement end-to-end functionality starting from an incoming client request, following into individual functions, across external services, ending with a synchronous response or after a chain of asynchronous event-based function triggers. %
    We implement realistic serverless applications and instrument them using distributed tracing to track end-to-end operation.
    \item \textbf{Heterogeneity:}
    A heterogeneous benchmark suite should include diverse applications by different dimensions. %
    Beyond including the most popular choice (e.g., most prevalent programming language), we strive for generalizability by covering additional applications.
    \item \textbf{Reproducibility:}
    A reproducible benchmark suite mitigates threats to internal validity that could affect the ability to obtain the same results with the same method under \emph{changed conditions} of measurements~\cite{taylor:94}.
    We provide automated containerized benchmark orchestration for all applications including their configurations and pinned dependencies.
    \item \textbf{Extensibility:}
    An extensible suite allows for adding existing serverless applications written in any programming language, using any framework, or cloud service dependencies with no or only minor code changes.
    We demonstrate the extensibility of our plugin-based harness by integrating existing applications maintaining their diverse structure rather than inventing new applications.
\end{enumerate}

\subsection{High-Level Design of ServiBench}
\emph{ServiBench (sb)} uses a data-driven benchmarking method based on a suite of serverless applications described in Section~\ref{sec:serverless_apps} and serverless invocation patterns derived from real traces as described in Section~\ref{sec:invocation_patterns}.

\begin{figure*}[htb]
    \centering
    \includegraphics[width=0.85\textwidth]{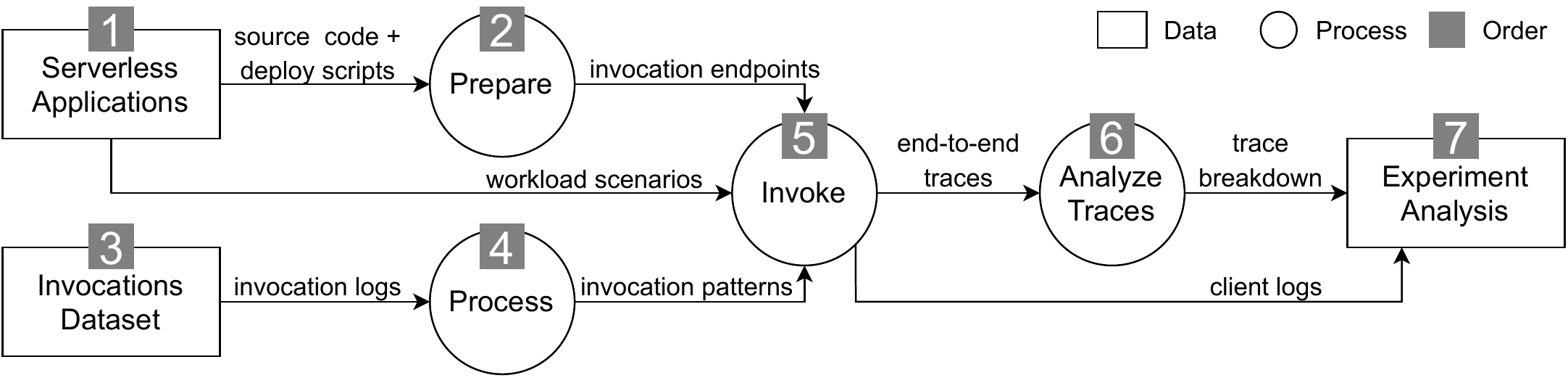}
    \vspace{-8pt}\caption{Data flow of a benchmark execution in sb.}\label{fig:sb_dataflow}
\end{figure*}

Figure~\ref{fig:sb_dataflow} illustrates the data flow and main processes of a benchmark execution with sb.
Each serverless application provides two additional assets in addition to its source code (Figure~\ref{fig:sb_dataflow}, label~\boxed{1}).
First, it defines deployment instructions annotated with OCI-compatible container images for dependency-bundled deployment.
This packaging enables reproducible cross-platform builds of the application and its service dependencies, which are defined through infrastructure-as-code~\cite{huttermann:12a}.
Second, a workload scenario defines how an application is invoked. This can be a single parameterized HTTP request or a probabilistic state-machine emulating users flow through a multi-action user story.
The \emph{prepare} process builds and deploys the application into a serverless platform and performs any preparations for benchmarking such as uploading datasets or extracting service endpoints~(\boxed{2}). %

The invocation patterns for benchmarking are derived from an invocations dataset~(\boxed{3}). %
The \emph{process} step takes invocation logs and upscales these to finer-grained (second-level) invocation frequencies, following different traffic shape patterns~(\boxed{4}). %
Subsequently, the \emph{invoke} process combines application-specific workload scenarios with generic serverless invocation patterns to invoke the application invocation endpoints with realistic serverless traffic patterns~(\boxed{5}). %
This workload generation yields client logs, which are mainly used for validation during experiment analysis as described in Section~\ref{sec:results}.
The main results are detailed end-to-end (e2e) traces collected from individual application components and correlated by a trace identifier passed along inter-service calls.
The \emph{analyze traces} process~(\boxed{6}) %
uses critical path analysis to perform latency breakdown analysis (see Section~\ref{sec:trace_analysis}), which provides a trace breakdown summary to the experiment analysis~(\boxed{7}). %

We implement \emph{sb} as a Python package that offers a CLI and SDK to orchestrate serverless application benchmarking.
To integrate with sb, an application needs to provide a Python file with three lifecycle methods to prepare, invoke, and cleanup itself.
Sb supports Docker to package application-specific build and deployment dependencies, and automatically manages directory mounts and provider credentials.

We use the Azure Function Traces~\cite{shahrad:20} as the invocations dataset and the 
k6~\cite{url:k6}
load testing tool as the invoker.
Though sb automatically configures the trace-based invocation patterns, application-specific workload scenarios are written in JavaScript.
For AWS, applications have been instrumented with
X-Ray~\cite{url:Amazon:XRay}
and the trace analyzer works with AWS services supported by X-Ray.
Other providers are currently partially supported, with two applications integrated with Azure and basic deployment infrastructure for IBM Cloud and Google Cloud. Expanding support for these providers is ongoing work. 

\subsection{Serverless Applications}\label{sec:serverless_apps}

Table~\ref{tab:app_list} characterizes the 10 serverless applications in the benchmark suite.
We select multiple applications for each of the most common types identified by survey studies~\cite{eismann:21-tse,williams:18} except for the type of \emph{operations and monitoring} because such applications are difficult to test in isolation.
In particular,
\emph{API} refers to synchronously invoked web endpoints (e.g., REST, GraphQL),
\emph{async} processing applications are triggered through events (e.g., an upload to a storage bucket triggers a function),
and \emph{batch} refers to larger computation tasks often simultaneously processed by multiple functions.

For the dominant serverless programming languages JavaScript (JS) and Python~\cite{leitner:19,eismann:21-tse}, we select multiple applications.
The popular enterprise languages Java, C\#, and increasingly Go are represented by one application each.

Our applications are of representative size, as datasets~\cite{shahrad:20} and surveys~\cite{leitner:19,eismann:21-tse} show that most applications are composed of 10 or fewer functions.

We cover the most popular external services used in serverless applications~\cite{leitner:19,shahrad:20,eismann:21-tse} with a focus on API gateways, persistency services (e.g., S3 cloud storage, DynamoDB cloud database), and cloud orchestration (e.g., AWS Step Functions).
Appendix~\ref{sec:app_description} describes and motivates each application.

\begin{table}
\footnotesize
\centering
\begin{tabular}{@{}lllrl@{}}
\toprule
\textbf{Application}  & \textbf{Type}           & \textbf{Lang.} & \textbf{\#} & \textbf{External Services} \\ \midrule
\boxed{A}~Minimal Baseline~\cite{url:AppA} & API & JS & 1 & $\Diamond$\quad\quad\quad\quad\quad\quad\quad \\
\boxed{B}~Thumbnail Gen.~\cite{yussupov:19a}    & Async & Java       & 2         & $\Diamond\bigotimes$\quad\quad\quad\quad\quad\quad\\
\boxed{C}~Event Processing~\cite{yussupov:19a}        & Async & JS & 8         & $\Diamond$\quad$\triangle\blacksquare$\quad\quad$\bigodot$ \\
\boxed{D}~Facial Recognition~\cite{url:AppD}    & Async & JS & 6         & $\Diamond\bigotimes$\quad\quad$\bigstar\spadesuit$\quad\\
\boxed{E}~Model Training~\cite{kim:19}       & Async & Python     & 1         & $\Diamond\bigotimes$\quad\quad\quad\quad\quad\quad\\
\boxed{F}~Realworld Backend~\cite{url:AppF}     & API                     & JS & 21        & $\Diamond$\quad\quad\quad\quad\quad$\bigodot$\quad\\
\boxed{G}~Hello Retail!~\cite{url:AppG}         & API                     & JS & 10        & $\Diamond\bigotimes$\quad\quad$\bigstar$\quad\hspace{-0.1mm}$\bigodot\square$ \\
\boxed{H}~Todo API~\cite{yussupov:19a}              & API                     & Go         & 5         &$\Diamond$\quad\quad\quad\quad\quad$\bigodot$\quad \\
\boxed{I}~Matrix Multipl.~\cite{yussupov:19a}  & Batch             & C\#         & 6         & $\Diamond\bigotimes$\quad\quad$\bigstar$\quad\quad\quad\\
\boxed{J}~Video Processing~\cite{kim:19}      & Batch             & Python     & 1         & $\Diamond\bigotimes$\quad\quad\quad\quad\quad\quad\\ \bottomrule
\end{tabular}
\vspace{-8pt}\caption{\label{tab:app_list} Characteristics of each end-to-end serverless application. Abbreviations: Programming language (Lang), number of functions (\#), API gateway ($\Diamond$), cloud storage ($\bigotimes$), cloud pub/sub ($\triangle$), cloud queue ($\blacksquare$), cloud orchestration ($\bigstar$), cloud ML ($\spadesuit$), cloud DB ($\bigodot$), cloud streaming ($\square$).}
\end{table}

\subsection{Serverless Invocation Patterns}\label{sec:invocation_patterns}

This section describes how we derive invocation patterns from the Azure Function Traces~\cite{shahrad:20}, a dataset invocation logs from a commercial cloud platform over two weeks.

\subsubsection{Selection and Classification} %
From the 74,347 functions in the Azure traces, we selected 528 functions with relevant properties for benchmarking  
logs~\cite{url:data:Azure}.
We removed 45,564 temporary functions not available over the entire two-week period and skip 15,319 timer triggers because these follow predictable periodic patterns and are typically not latency-critical.
Knowing that the 18.6\% most popular applications with invocation rates $\geq 1/min$ represent 99.6\% of all function invocations~\cite{shahrad:20}, we selected the 2.6\% most popular functions with average invocation rates $\geq 1/s$ 
as they are relevant for high-volume benchmarking. %

\begin{figure}[t]
    \centering
    \vspace{-12pt}\includegraphics[width=0.48\textwidth]{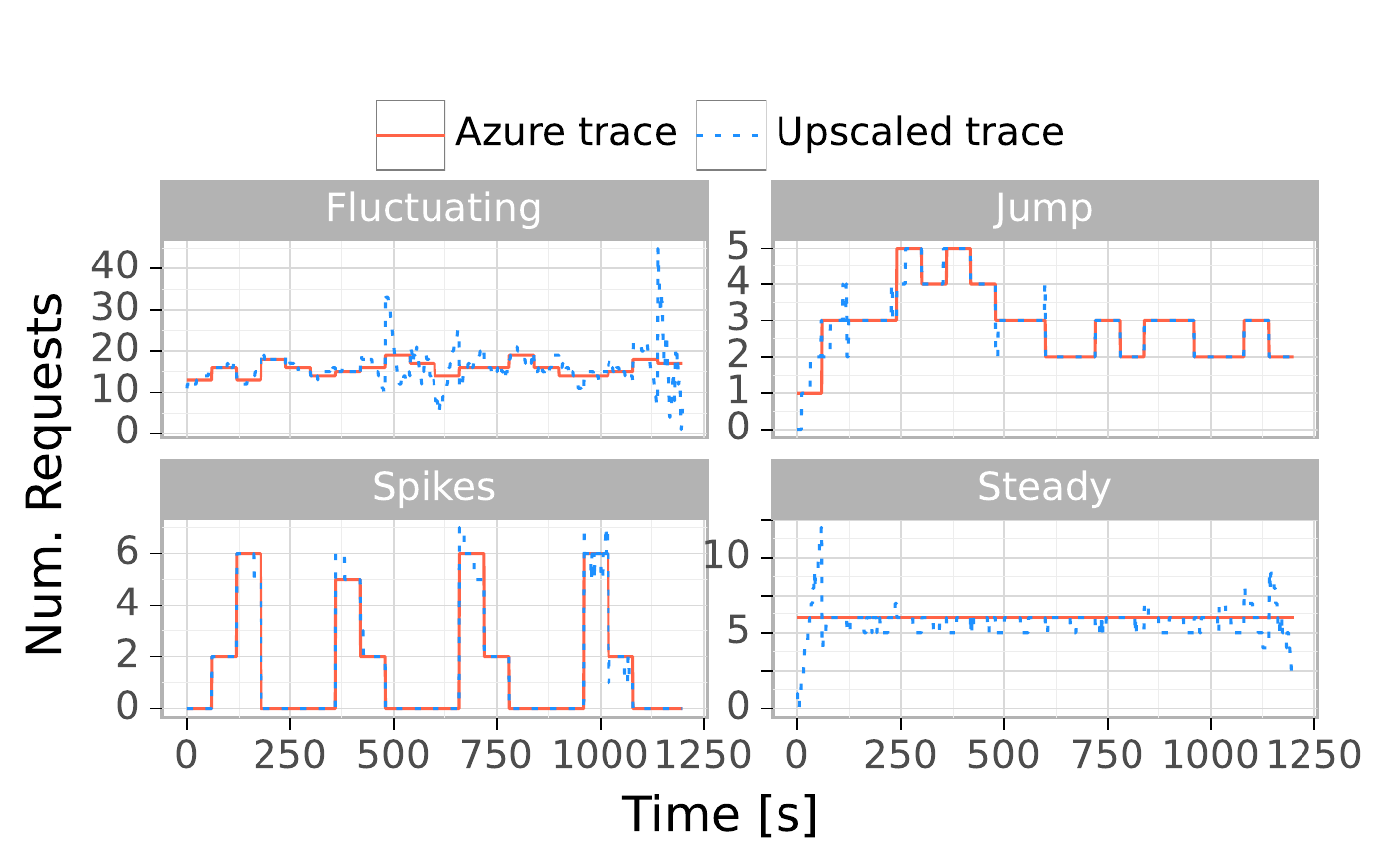}
    \vspace*{-0.9cm}
    \caption{Typical serverless invocation patterns over 20 min.}\label{fig:invocation_patterns}
\end{figure}

We visually identified 4 typical invocation patterns by manually classifying two time ranges for 100 of the selected 528 functions.
We first created 200 individual line plots with invocation counts over 20 minutes\footnote{We explored different time resolutions (2 weeks, 1 day, 4 hours, 1 hour, 30 min, 20 min, 10 min) and found that hourly patterns are similar enough to 20 minutes, which is feasible cost-wise for repeated experimentation with many different applications under varying configurations.} and grouped similar traffic shapes into several clusters.
After merging similar patterns, we identified 4 common patterns (see Figure~\ref{fig:invocation_patterns}):
\begin{enumerate*} [label=\upshape(\roman*\upshape)]
    \item \emph{steady}~(32.5\%) represents stable load with low burstiness,
    \item \emph{fluctuating}~(37.5\%) combines a steady base load with continuous load fluctuations especially characterized by short bursts,
    \item \emph{spikes}~(22.5\%) represents occasional extreme load bursts with or without a steady base load, and
    \item \emph{jump}~(7.5\%) represents sudden load changes maintained for several minutes before potentially returning to a steady base load.
\end{enumerate*}

\subsubsection{Trace Upscaler}
\label{sec:trace_upscaler}

The trace upscaler generates invocation rates at the granularity of seconds from per-minute invocation logs.
The Azure dataset~\cite{shahrad:20} reports the number of function invocations per minute.
However, bursty serverless invocation rates are not uniformly distributed over a minute~\cite{wang:21a}.
For example, it could be that the majority of the invocations in a single minute all occur in the first five seconds.

To generate more realistic patterns than uniformly distributed or linearly interpolated invocation rates, we use fractional Brownian motion to synthesize perturbations at the granularity of seconds while maintaining the large scale properties of the Azure traces.
We chose a Hurst parameter of 0.8 as empirically determined for web traffic~\cite{crovella:97}.
This indicates positive correlation over time, which means an increase in the workload is likely followed by an increase and a decrease is likely followed by a decrease.
We believe this to be realistic workload as it generates patterns where workload bursts are sustained instead of oscillating between peaks and valleys.

\section{Detailed Distributed Trace Analysis for Serverless Architectures}\label{sec:trace_analysis} %
\begin{figure*}[htb]
    \centering
    \includegraphics[width=\linewidth]{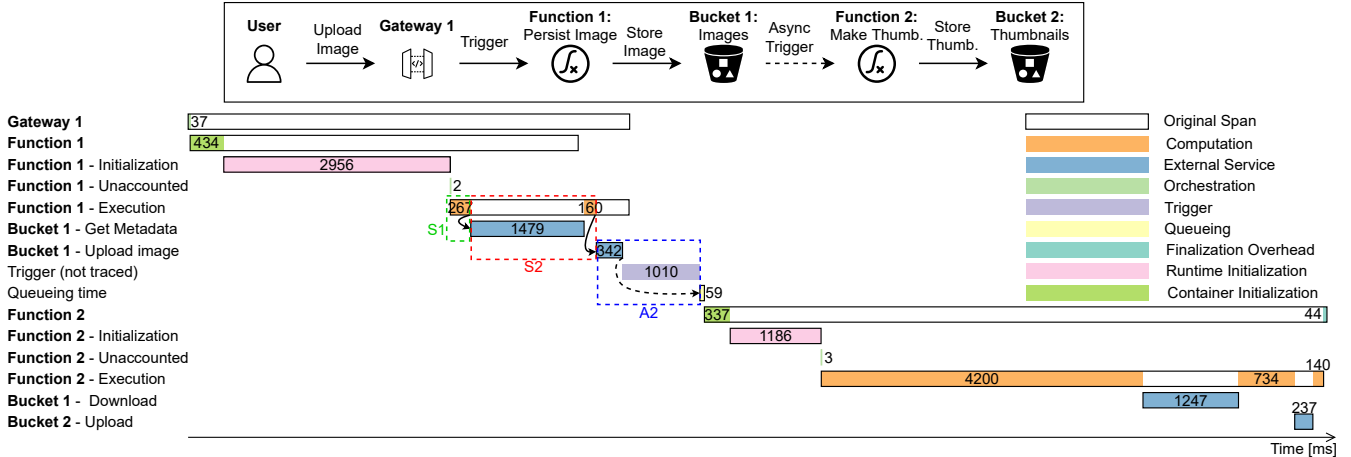}
    \vspace*{-0.9cm}
    \caption{Simplified depiction of an \emph{execution trace} (\definition{1}) with annotated \emph{latency breakdown} (\definition{3}). \\Data collected from \app{A} with two cold starts. Values represent time in milliseconds. 
    Labels Sync/Async refer to  Figure~\ref{fig:breakdown_cases}.}\label{fig:trace_breakdown}
\end{figure*}

We motivate distributed tracing of serverless applications and describe how to extract the critical path and a detailed latency breakdown from distributed traces.
\subsection{Challenges and Background}
Distributed tracing has been adopted for various use cases~\cite{mace:17,sambasivan:14} such as distributed profiling (i.e., latency analysis), anomaly detection (i.e., identifying and debugging rare problems), or workload modeling (e.g., identifying representative workflows).
Tracing systems such as Google's Dapper~\cite{sigelman:10} or Facebook's Maelstrom~\cite{veeraraghavan:18} help improve performance, correctness, understanding, testing, and recovery of services.
These insights are even more important for highly distributed serverless architectures given their ephemeral nature.
However, event-based coordination is inherently asynchronous, hence hard-to-track background workflows need to be included in the tracing and cannot be ignored as for synchronous microservice architectures~\cite{qiu:20}.
The limited control in serverless environments makes users dependent on provider tracing implementations or resort to less detailed third-party or custom implementations.
Further, tracing issues are common at large scale and trace analysis must detect and handle clock inaccuracy and incomplete traces.

We represent each request as an \emph{execution trace} where the \emph{critical path} determines the end-to-end latency and the \emph{latency breakdown} lists and classifies each time span along the critical path as visualized in Figure~\ref{fig:trace_breakdown}.

\paragraph{Definition~1}\label{def:1}
An \emph{execution trace} of a serverless application is a causal-time diagram of the distributed execution of a request, where a node is a \emph{trace span} that corresponds to an individual unit of work (e.g., computation) and an edge represents a causal relationship through a synchronous or asynchronous invocation.
Each trace span contains a start and end timestamp and is correlated by a trace id.
\paragraph{Definition~2}\label{def:2}
A \emph{critical path} in an execution trace is the longest path weighted by duration, which starts with a client request and ends with the trace span that has the latest end time.
This definition of end-to-end latency includes asynchronous background workflows that do not return to their parent spans to capture the event-based nature of serverless systems.
Hence, our definition differs from a critical path of a synchronous client response in microservices~\cite{qiu:20}.

\paragraph{Definition~3}\label{def:3}
A \emph{latency breakdown} of an execution trace is the most detailed list of time segments along the critical path without any temporal gaps.
This explicitly includes transitions between trace spans, which are often implicit in an execution trace.
In its aggregated form\footnote{We use high-level categories for better readability and cross-application comparison as the full trace breakdown is very detailed. Individual external services, such as cloud storage, could be classified separately if needed.}, each time segment is classified and summed up by the following categories common to serverless applications:
\begin{enumerate*} [label=\upshape(\roman*\upshape)]
    \item \emph{computation} represents the actual processing time of serverless functions.
    \item \emph{external service} represents the time waiting for the completion of a services request (e.g., database query, file upload to a storage services).
    \item \emph{orchestration} represents time spent coordinating serverless function executions by workflow engines (e.g., AWS Step Functions) or API gateways dispatching requests to functions.
    \item \emph{trigger} represents the implicit transition time between an event and a function bound to this event (e.g., time between enqueuing a message until the event is dispatched to a function).
    \item \emph{queuing} represents the time spent in function worker queues before it starts executing.
    \item \emph{container initialization} represents the time it takes to provision the function execution environment.
    \item \emph{runtime initialization} represents the time it takes to initialize the function language runtime during a cold start.
    \item \emph{finalization overhead} represents cleanup tasks after function execution and before freezing the sandbox.
\end{enumerate*}

\subsection{Latency Breakdown Extraction}
We first extract the \emph{critical path} of an \emph{execution trace} and subsequently refine it into a detailed \emph{latency breakdown}.

To extract the critical path, we use Algorithm~\ref{alg:critical_path}, which is a modified version of the weighted longest path algorithm proposed in the context of microservices~\cite{qiu:20}.
Our modifications support asynchronous invocations, unordered traces, and refined heuristics to handle timing issues~\cite{najafi:21} in fine-grained serverless tracing.
A stack with all parent spans of the last ending span is used to only recurse into child spans connected to the last ending span.
Line~9 defines the sorting order for child spans primarily by the \emph{endTime} and secondarily by the \emph{startTime}.
The secondary sort key is required to handle the special case of a single trace span with a duration of 0 milliseconds.  %
Our heuristic \textsc{happensBefore} detects sequential relationships and \textsc{isAsync} detects asynchronous invocations.
They support a configurable error margin (default 1ms) to gracefully handle minor clock inaccuracies by implementing temporal comparisons such as $t_{1} < t_{2}$ with $t_{2} - t_{1} + margin$.

\begin{algorithm}[tb]
\caption{Critical Path Extraction, based on~\cite{qiu:20}.}\label{alg:critical_path}
\begin{algorithmic}[1]
\Require Serverless execution trace $T$ with\newline
$span$ attributes $childSpans$, $startTime$, $endTime$ and\newline
stack $S$ with all parent spans of the last ending span
\Procedure {$T$.\textsc{CriticalPath}}{$S$, $currentSpan$}{}
\State $path \leftarrow [$currentSpan$]$
\If {$S$.top() == $currentSpan$}
    \State $S$.pop()
\EndIf
\If {$currentSpan.childSpans == None$}
    \State Return $path$
\EndIf
\State $sortedChildSpans \leftarrow$ sortAscending(\newline
$currentSpan.childSpans$, by=[$endTime$, $startTime$])
\State $lastChild \leftarrow sortedChildSpans.last$
\For {each $child$ in $sortedChildSpans$}
    \If {$child$.\textsc{happensBefore}($lastChild$) and\newline
        not $currentSpan$.\textsc{isAsync}($path$.last)}
        \State $path$.extend($\textsc{CriticalPath}(S, child)$)
    \EndIf
\EndFor
\If {($currentSpan$.\textsc{isAsync}($lastChild$) and\newline
$S$.top() == $lastChild$) or\newline
(not $currentSpan$.\textsc{isAsync}($lastChild$) and\newline
not $currentSpan$.\textsc{isAsync}($path$.last)}
    \State $path$.extend($\textsc{CriticalPath}(S, lastChild)$)
\EndIf
\State Return $path$
\EndProcedure
\Statex
\Procedure{$current$.\textsc{happensBefore}}{$next$}{}
    \State Return $current.endTime < next.startTime$
\EndProcedure
\Procedure{$current$.\textsc{isAsync}}{$next$}{}
    \State Return $next.endTime > current.endTime$
\EndProcedure
\end{algorithmic}
\end{algorithm}

\begin{figure}[tb]
    \centering
    \includegraphics[width=0.45\textwidth]{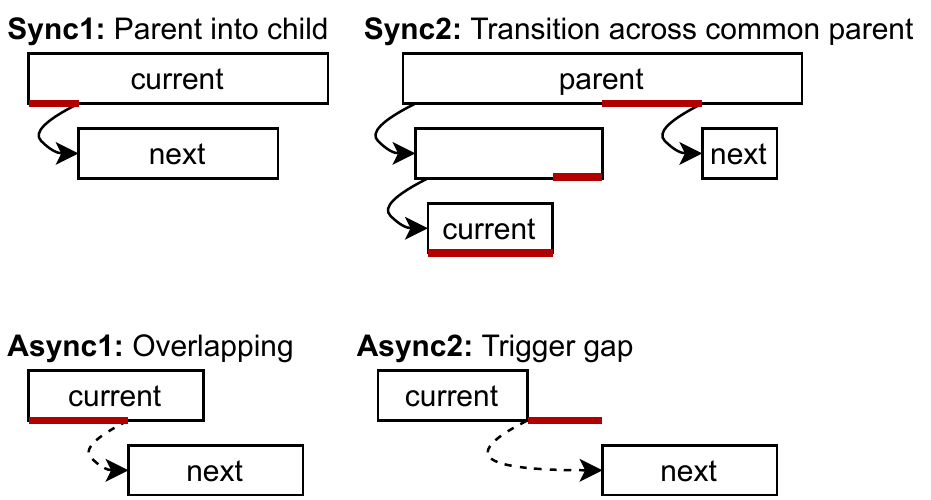}
    \vspace{-4pt}\caption{Extraction cases of latency breakdown~(red segments) for pairs of current and next nodes on the critical path.}\label{fig:breakdown_cases}
\end{figure}

We extract the detailed latency breakdown along the critical path by identifying and categorizing every time segment while accounting for all gaps between spans.
Figure~\ref{fig:breakdown_cases} visualizes the common cases for synchronous and asynchronous invocations while iterating pairwise (\emph{current}, \emph{next}) over the critical path.
For synchronous invocations, we distinguish two different cases:
\emph{Sync1} handles a traditional synchronous invocation from a \emph{current} parent span into a \emph{next} child span.
\emph{Sync2} handles a potentially recursive transition from the \emph{current} span on a synchronous invocation stack across a common \emph{parent} into its \emph{next} child span.
For asynchronous invocations, we distinguish two cases:
\emph{Async1} handles if the \emph{next} child span overlaps with the \emph{current} parent span.
\emph{Async2} handles if there is a gap between the \emph{current} parent span and the \emph{next} child span.
This scenario frequently occurs in serverless systems when triggering a function using a slow trigger.
There is a third case that we can currently not detect, which is structurally equivalent to \emph{Sync1} except that the call to \emph{next} is asynchronous.
To make this case detectable, trace specifications could define labels for synchronous and asynchronous parent-child relationships as discussed for Open Telemetry~\cite{url:tool:OpenTelemetry}.
Finally, we assign an activity label (e.g., computation) to each breakdown segment depending on the span type (e.g., function) as annotated in Figure~\ref{fig:trace_breakdown}.

\section{Experimental Results}\label{sec:results} %

We use in this section ServiBench to comprehensively benchmark the performance of a popular serverless platform.
We deploy ServiBench on the serverless platform AWS Lambda, which various reports~\cite{sldataset,eismann:21-tse} indicate is much used by serverless applications in production.

ServiBench supports many real-world performance scenarios, from which we focus in this work on 
(1) latency breakdown to understand the performance of warm invocations and of cold starts, and on (2) the impact of invocation patterns on (median) end-to-end latency.
We make 6 observations, and discuss their implications for serverless practitioners and researchers in Section~\ref{sec:exp:discussion}.

\subsection{Experiment Design} %

We conduct a performance benchmarking experiment~\cite{hasselbring:21} with an open-loop load generator in the data center region Northern Virginia (us-east-1) as commonly used by other serverless studies~\cite{wang:18,cordingly:20a,copik:21a,barcelona-pons:21,wen:21b}.
We collected over 7.5 million traces, through over 12 months of experimentation in 2021 and 2022.
\paragraph{Application configuration}
All functions are configured with the same memory size of 1,024\,MB as this provides a balanced cost-performance ratio~\cite{eismann:21a} between the minimal memory size of 128\,MB (heavy CPU throttling) and the maximum memory size for a single CPU core of 
1,769\,MB\cite{url:data:Lambda}
(inefficient for non-CPU-intensive load).
For application-specific memory size tuning, we refer to
\emph{aws-lambda-power-tuning}~\cite{url:tool:AWSLPT},
systematic literature reviews~\cite{raza:21,scheuner:20-jss}, and many empirical studies~\cite{akhtar:20,lloyd:18a,wang:18,figiela:18,wang:19,yu:20,eismann:21a}.
All supported cloud services (API Gateway, Lambda, Step Functions) have distributed X-Ray tracing enabled to trace every request.
For applications with multiple endpoints, we present one representative endpoint in the paper and refer to the replication package for detailed results.

\begin{figure}[tb]
    \centering
    \includegraphics[width=0.47\textwidth]{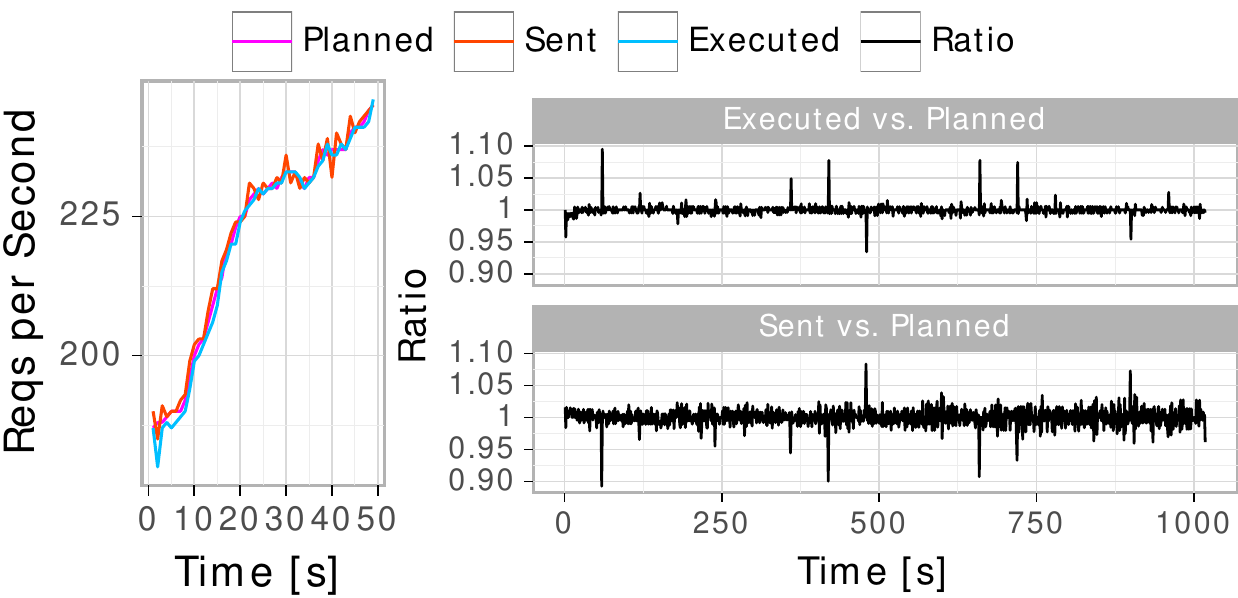}
    \vspace{-6pt}\caption{Comparison of per-second invocation rates planned vs. sent vs. executed (left) and validation ratio for pairwise comparison (right).}
    \label{fig:load_validation}
\end{figure}

\paragraph{Load generator}
For accurate load generation, we deploy an over-provisioned EC2 instance of the type \emph{t3a.large} in the same region as the serverless applications.
We validate per-second invocation rates for accurate load generation (planned vs. sent) by correlating the load configuration with the client logs and actual load serving (generated vs. executed) by correlating the client logs with the backend traces.
We combine visual comparison (see Figure~\ref{fig:load_validation}) with FastDTW~\cite{salvador:07}, an approximate Dynamic Time Warping (DTW) algorithm.
We monitor application error rates client-side by checking response status codes and server-side by checking for any exceptions in each trace.
Finally, we investigate any invalid traces due incomplete or invalid trace data following both logical and time-based validation.
\subsection{Latency Breakdown}\label{sec:latency_breakdown}
We first drill down into the end-to-end latency of serverless applications to identify critical components using sb (Section~\ref{sec:benchmark_suite}) and trace breakdown extraction (Section~\ref{sec:trace_analysis}).
This application-level perspective complements existing work, which primarily focused on micro-benchmarking individual components or reporting client-side response times for synchronously orchestrated applications~\cite{scheuner:20-jss,raza:21}.
As a baseline, we focus on warm invocations and subsequently compare the latency penalty of cold invocations and tail latency.

\paragraph{Method.}
For each of the 10 applications from Section~\ref{sec:app_description}, we send 4 bursts of 20 concurrent requests with an inter-arrival time of 60 seconds between each burst.
The first burst triggers up to 20 cold invocations used in Section~\ref{sec:coldstarts} and after the function completes within 60 seconds, the following 3 bursts trigger more warm invocations used for tail-latency analysis in Section~\ref{sec:tail_latency} and as baseline in Section~\ref{sec:warm_invocations}.
To collect enough samples under the same conditions, we conduct 10~trials and 14~repetitions resulting in up to 8,400~warm invocations (3 $\times$ 20 $\times$ 10 $\times$ 14)\footnote{A related study~\cite{ustiugov:21a} uses 3,000~samples for individual functions; we target more per-application samples as requests can be distributed across endpoints.}.
For each of the 10 trials, we invoke each application using round-robin scheduling with inter-trial times of 50 minutes to trigger cold invocations in the first burst for Section~\ref{sec:coldstarts}.
Before each of the 14 repetitions of trials, we re-deploy each application to ensure a clean state. %
We use trace analysis to detect cold invocations through the presence of
\emph{Initialization} segments~\cite{url:Amazon:XRay:using}.
For applications with chained functions, we ignore ``partial cold starts'' and only consider ``full'' warm or cold invocations, where every function in the critical path shares the same cold start status.

\subsubsection{Warm Invocations}\label{sec:warm_invocations}
Frequently invoked applications often get warm invocations.

\begin{figure}[!t]
    \centering
    \includegraphics[width=0.45\textwidth]{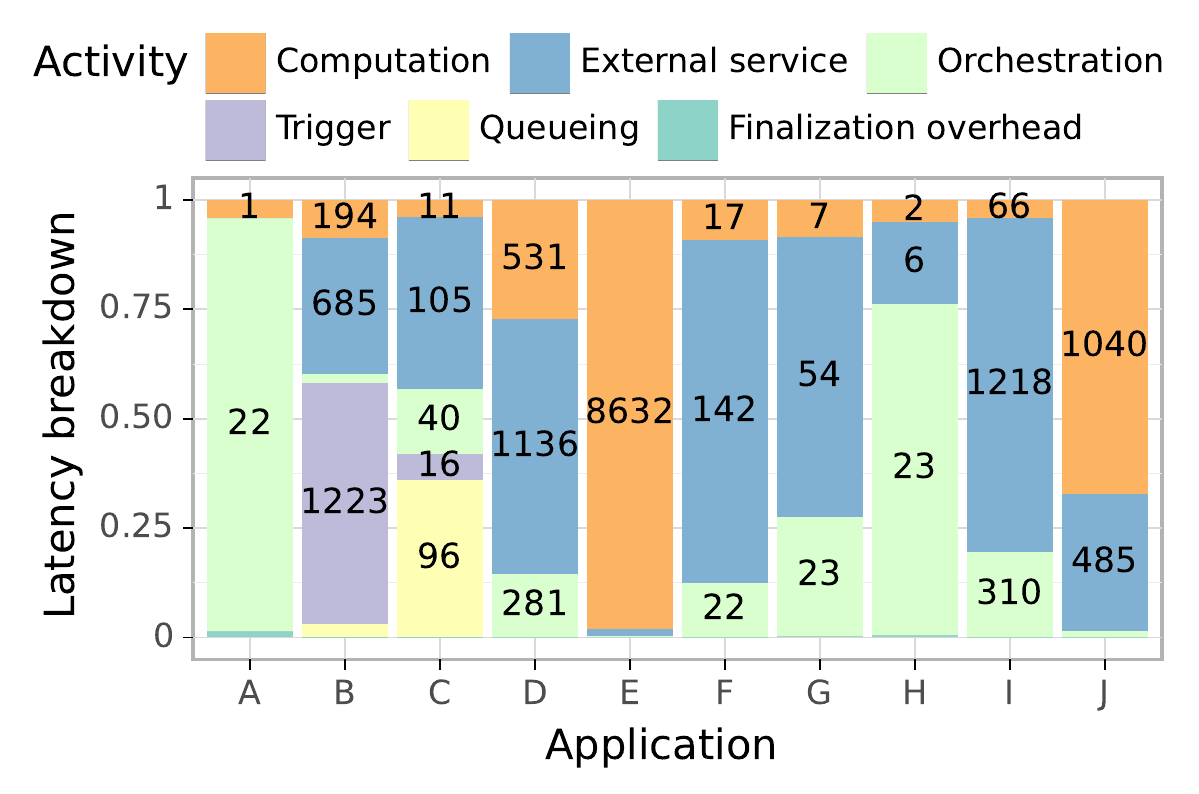}
    \caption{Latency breakdown of warm invocations as median fraction of end-to-end latency. 
    Values inside the bar-stacks represent absolute time per activity, in milliseconds.}\label{fig:latency_warm_p50}
\end{figure}

\paragraph{Results.}
The relative latency breakdown in Figure~\ref{fig:latency_warm_p50} shows the median latency for each activity introduced in \definition{3}.
\app{A} exemplifies the orchestration overhead~(22\,ms) of a common serverless pattern where an API gateway is connected to a function.
Lightweight applications such as \app{H} are similarly dominated by \emph{orchestration} time~(23\,ms) because they do minimal computation work and use fast external services (e.g., 6\,ms database insert).
\app{E} and \app{J} are examples of computation-heavy workloads.
External services like blob storage or computer vision APIs are often the dominating factor, especially for many I/O operations~(\app{I}) or larger files~(\app{D}).
Asynchronous applications are typically dominated by transition delays due to trigger and queuing time as demonstrated by the applications \app{B} and \app{C}.

\begin{observationbox}\textbf{Observation 1:} The median end-to-end latency of serverless applications is often dominated by external service calls and synchronous orchestration or asynchronous trigger-based coordination.
The actual computation time in serverless functions is relatively little except for inherently compute-heavy workloads.
\end{observationbox}

\subsubsection{Cold Starts}\label{sec:coldstarts}

We now study which time categories contribute to higher cold start latency  using the results from Section~\ref{sec:latency_breakdown} as a baseline.
Tracing cold starts requires access to timestamps captured within provider-internal infrastructure.
Sb can extract these internal timestamps from AWS X-Ray traces and distinguish between container and runtime initialization time, which would only be possible for self-hosted~\cite{ustiugov:21} or provider-internal~\cite{agache:20,brooker:21} systems otherwise.
Insights on cold starts are relevant for applications that are invoked irregularly~(e.g., inter-arrival times > 10 minutes) or exhibit bursty invocation patterns and, hence, need to provision new function instances.
\begin{figure}[!t]
    \centering
    \includegraphics[width=0.45\textwidth]{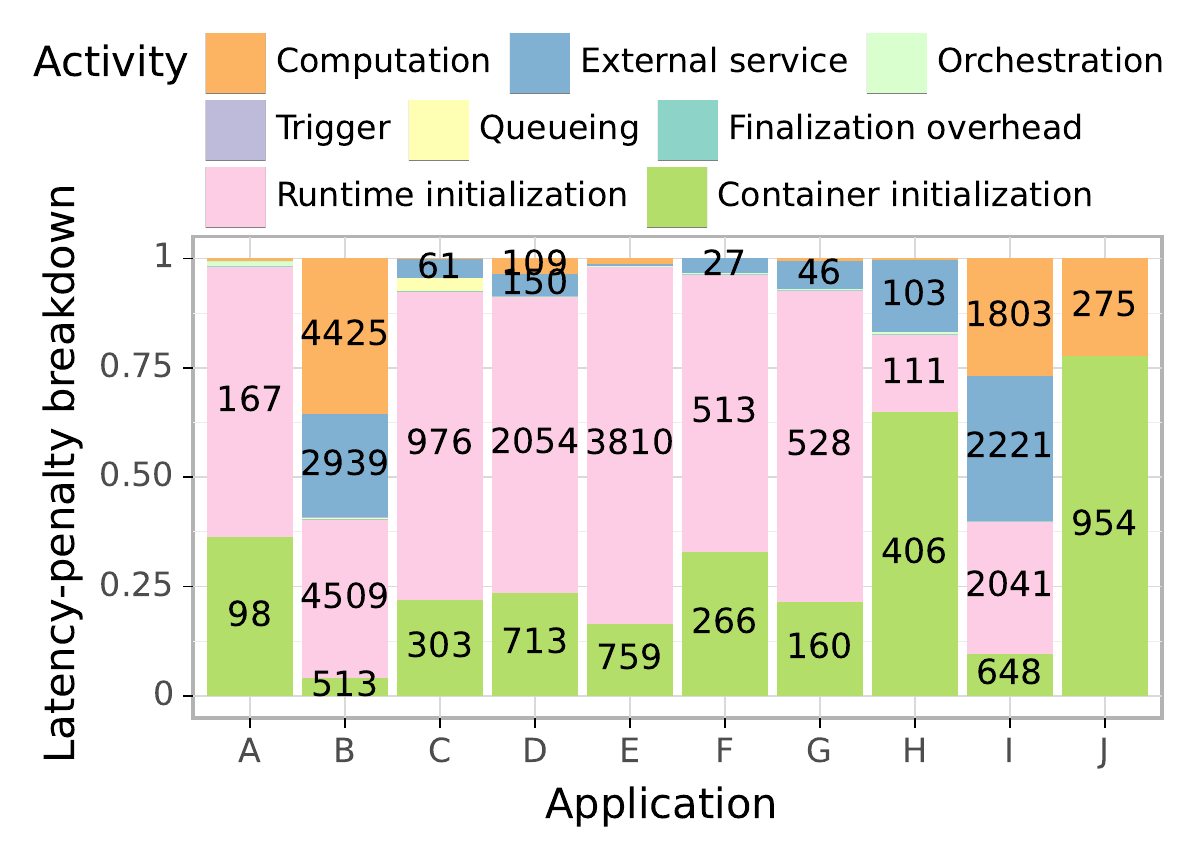}
    \caption{Latency-penalty breakdown for cold invocations compared to the baseline of warm invocations~(Figure~\ref{fig:latency_warm_p50}) as fraction of the difference between the medians. %
    Values inside the bar-stacks are absolute, in milliseconds~(ms), e.g., for \app{B}, Computation 
    takes 4,425\,ms longer.}\label{fig:latency_warm_cold_diff_p50}
\end{figure}

\paragraph{Results.}
Figure~\ref{fig:latency_warm_cold_diff_p50} shows the latency difference between the medians for cold invocations compared to warm invocations.
\app{A} depicts a common initialization overheads for a function behind an API gateway of 265\,ms~(98+167) in line with prior cold start studies for Node.js by Wang et al.~\cite[Figure~6]{wang:18} and Maissen et al.~\cite[Figure~7]{maissen:20}.
Our results are more detailed and reveal differences for realistic applications.
Our trace details show that runtime initialization typically accounts for the majority of cold start overhead compared to container initialization.
For \app{A}, the container initialization time of 98\,ms is \textasciitilde 20\,ms faster than the boot times for the underlying Micro VMs as reported for pre-configured Firecracker~\cite[Figure~6]{agache:20}.
In comparison to other applications, the relative latency penalty remains similar~(c.f., \app{J}).
However, other realistic applications have much higher absolute initialization times due to large packaged dependencies and chains of multiple functions in the critical path.

Beyond runtime and container initialization, other categories can add cold start overhead that is often overlooked.
Computation can contain conditional code executed only upon cold starts or trigger one-off optimizations such as just-in-time compilation for 
interpreted languages~\cite{blog:Minic21}
exemplified by the applications \app{B} in Java and \app{I} in C\#.
External service time can add connection overhead due to extra authentication upon cold starts~(e.g., \app{B} caches S3 authentication) or database connection setup~(e.g., \app{H} connects to DynamoDB).
Finally, the following categories related to application coordination remain unaffected by cold starts: orchestration, trigger, queuing, and instrumentation overhead.

\begin{observationbox}\textbf{Observation 2:} Runtime initialization and container initialization add most overhead for cold invocations but external service connection initialization and one-off computation tasks can also contribute.
\end{observationbox}

\subsubsection{Tail Latency}\label{sec:tail_latency} %

Tail latency is increasingly important at scale for cloud providers~\cite{dean:13} and hence particularly challenging for massive multi-tenant serverless systems.
Prior studies~\cite{ustiugov:21a,pelle:19,wang:18,lee:18} conducted micro-benchmarks to measure tail of individual serverless components.
By leveraging our trace analysis~(Section~\ref{sec:trace_analysis}), we can directly identify which time categories contribute to tail latency~(99th percentile) for entire applications.
\begin{figure}[!t]
    \centering
    \includegraphics[width=0.43\textwidth]{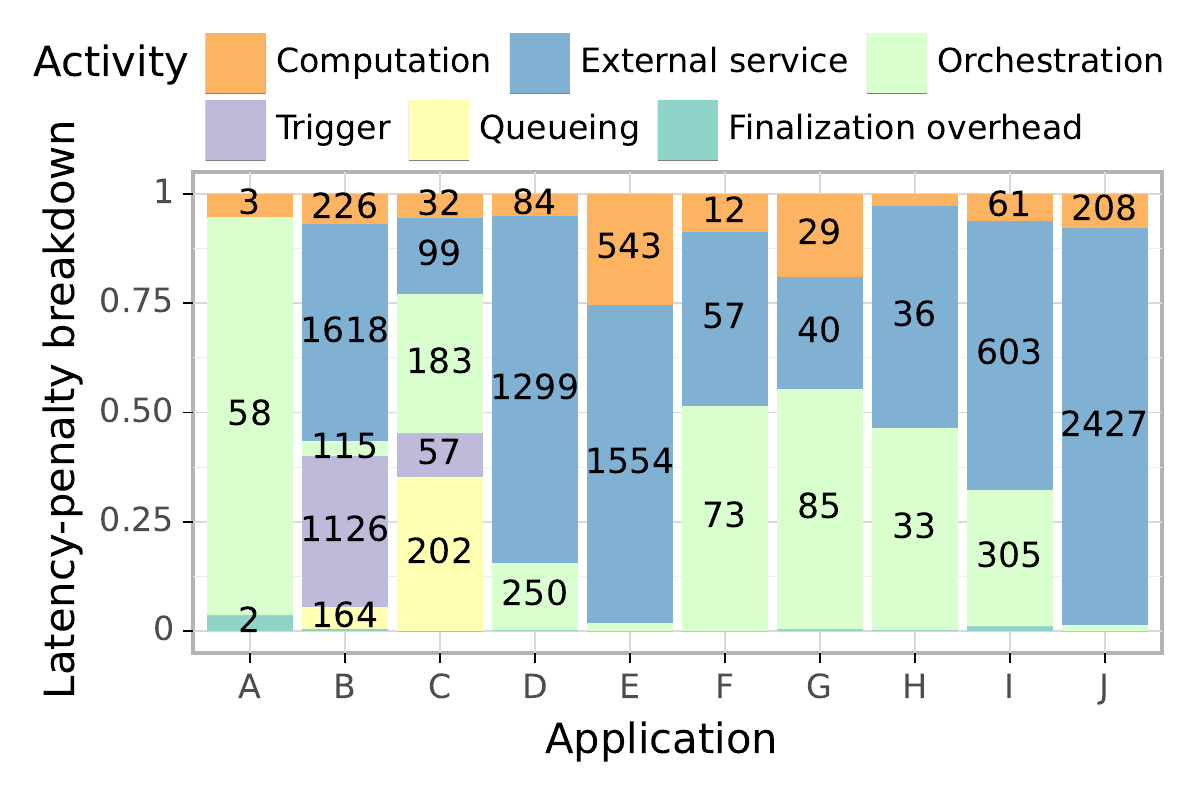}
    \vspace{-0.4cm}
    \caption{Latency-penalty breakdown for slow invocations compared to baseline of warm invocations~(Figure~\ref{fig:latency_warm_p50}) as fraction of the difference between the median and 99\textsuperscript{th} percentile. Values inside the bar-stacks are absolute, in milliseconds~(ms), e.g., for \app{J}, the Cloud storage service adds 2,427\,ms of delay for slow invocations; this accounts for \textasciitilde 90\% of the tail-latency slowdown.}\label{fig:slow_warm_requests}
\end{figure}

\paragraph{Results.}
Figure~\ref{fig:slow_warm_requests} shows that external services cause major variability. %
In particular, storing a large file~(+2,429\,ms for \app{J}) causes massively more tail-latency delay than storing many chunks of small files~(+602\,ms for \app{J}).
Database services contribute less to tail latency than object storage as demonstrated by the applications \app{C}, \app{F}, \app{G}, and \app{H} with latency penalties between 35\,ms and 99\,ms.

Another factor of tail latency is the serverless overhead for orchestrating synchronous applications (i.e., \emph{orchestration} time) and asynchronous applications (i.e., \emph{trigger} and \emph{queuing} time).
These categories double or triple their latency in comparison to the baseline in Figure~\ref{fig:latency_warm_p50}.
Computation is inherently variable in a multi-tenant system but contributes at most 25\% to the latency penalty for compute-heavy \app{E}.

\begin{observationbox}\textbf{Observation 3:} Tail latency is primarily caused by external services, particularly by object storage.
\end{observationbox}

\subsection{Invocation Patterns} %
\label{sec:res_invocation}
Real-world applications exhibit diverse invocation patterns~\cite{shahrad:20} but prior work rarely investigated dynamic workloads over time~\cite{lee:18} or different invocation patterns~\cite{kuhlenkamp:20} and if so, using artificial applications, patterns, and one-time bursts~\cite{ustiugov:21a,wang:18,barcelona-pons:21}.
It remains unclear how different invocation patterns derived from the Azure Function Traces~\cite{shahrad:20} (Section~\ref{sec:invocation_patterns}) affect the end-to-end latency of serverless applications.
To address this gap, we investigate the performance effect of varying invocation rates over time under an equivalent average invocation rate. %

\begin{table}[!t]
\footnotesize
\centering
\begin{tabular}{@{}cccccccccc@{}}
\toprule
\boxed{A} & \boxed{B} & \boxed{C} & \boxed{D} & \boxed{E} & \boxed{F} & \boxed{G} & \boxed{H} & \boxed{I} & \boxed{J} \\ \midrule
200     & 37     & 50     & 25     & 22     & 167     & 154     & 200     & 10     & 25     \\ \bottomrule
\end{tabular}
\caption{Invocation rates~(in reqs./s) used per application, set at 50\% of the achieved load in the scalability pre-study.}
\label{tab:load_levels}
\end{table}

\paragraph{Scalability prestudy.}
We conducted a prestudy to adjust the average invocation rates to our 10 heterogeneous applications.
Using the same invocation rate or concurrency level for all applications is inappropriate because it overloads some applications while others remain close to idle\footnote{We collected over 700K traces for \app{B} to \app{J} using the invocation patterns in Section~\ref{sec:invocation_patterns} with an average rate of 20~reqs/sec. We tried different concurrency levels, but noticed that long-running applications were overloaded and short-running applications were served by few function instances.}.
Therefore, we test increasing load levels with constant arrival rates for 90 seconds until an application exceeds a rate of 5\% for trace errors or invalid traces twice in succession.
Trace-based root cause analysis identified rate limits and function tracing issues\footnote{We reported this and additional issues related to clock drifting and trace correctness to AWS for further investigation.} as common. %
To avoid overloading an application, we select 50\% of the achieved load level as target average invocation rate for parameterizing the invocation patterns~(Table~\ref{tab:load_levels}).

\paragraph{Method.}
We treat each application from Table~\ref{tab:app_list} with 2 artificial and 4 realistic workloads derived from real-world traces as described in Section~\ref{sec:invocation_patterns}.
The artificial workloads serve as baseline for fully \emph{constant} load and maximal burstiness simulated by \emph{on\_off} alternations with load for 1 second and idle time of 3 seconds.
We scale the average invocation rate per-application following Table~\ref{tab:load_levels}.
We discard warmup measurements of the first 60 seconds as the actual invocation rate can deviate from the target rate in the first second and initial cold starts dominate the start of every experiment.
\begin{figure}[tb]
    \centering
    \includegraphics[width=0.49\textwidth]{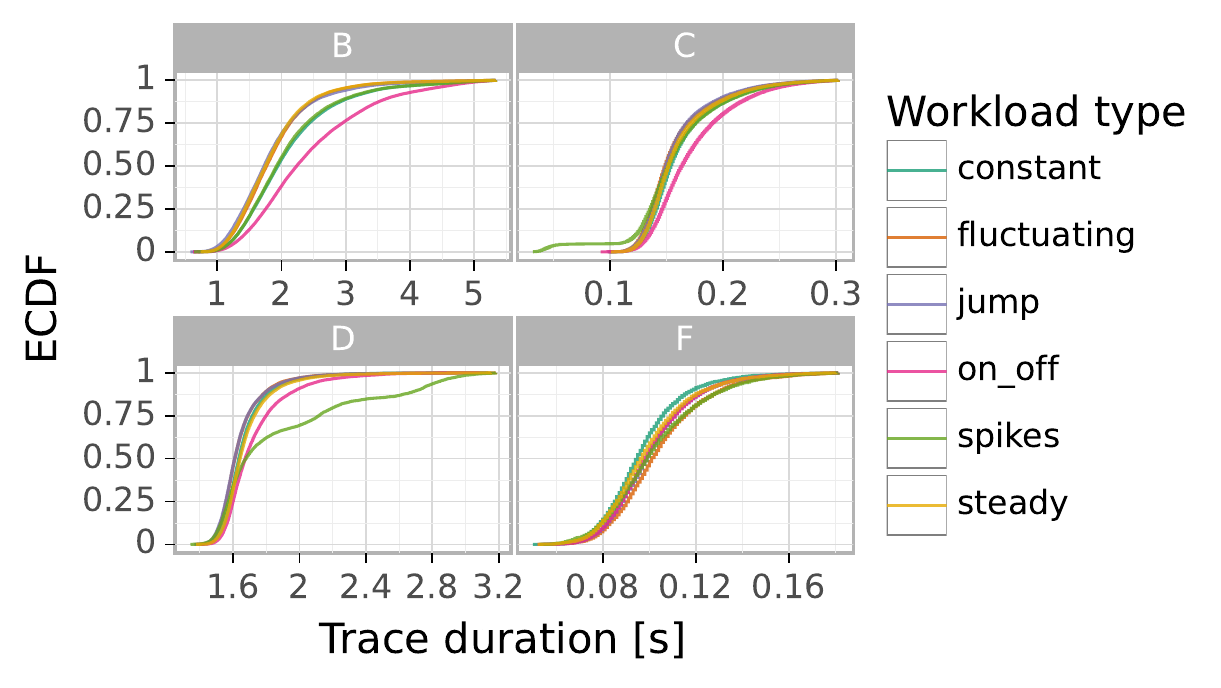}
    \vspace{-20pt}\caption{End-to-end latency for different invocation patterns clipped at the 99th percentile due to extreme long tail.}\label{fig:latency_by_workload_type}
\end{figure}

\paragraph{Results.}
Figure~\ref{fig:latency_by_workload_type} shows the partial CDF of the e2e latency for applications that accurately followed the target invocation pattern (<10\% deviation from target invocation rate and error metrics).
The median latency is unaffected by invocation patterns as shown by the overlapping CDF curves.
Percentiles up to p99 clipped in the CDF also show no relevant difference with the exception of \app{D}, where the peak invocation rates of the \emph{spikes} workload reach the rate limit of the facial recognition service causing external service delays.

The number of initial cold starts differs by invocation pattern but remains very low after the 60 seconds warmup time.
The \emph{on\_off} and \emph{spikes} patterns have higher peak invocation rates and trigger more initial cold starts.
However, after the warmup time, additional cold starts are rare~(below 10).
\begin{observationbox}\textbf{Observation 6:}
Different invocation patterns do not meaningfully affect the median end-to-end latency. %
\end{observationbox}

\subsection{Discussion} \label{sec:exp:discussion}

Our results emphasize the importance of serverless benchmarking that integrates fine-grained latency breakdown analysis, realistic invocation patterns, and varied benchmark applications.
Furthermore, our experiments lead to relevant implications for serverless practitioners and researchers.

\paragraph{Serverless fulfills its core promise of stable performance under bursty workloads.}
Our results show that serverless is indeed well-suited for bursty workloads after initial cold starts and when staying below platform-specific rate limits.
Hence, serverless fulfills its promise of built-in scalability under the given load levels for our 10 applications.
This result was somewhat unexpected, especially contrasting with prior research~\cite{ustiugov:21a,barcelona-pons:21,lee:18}.
However, of course, bursty workloads may still negatively impact performance on different platforms or with even more rapid bursts than what we evaluated (e.g., per-microsecond bursts rather than per-second bursts).

\paragraph{Slowdowns are caused by control flow and coordination, not computation.}
Our results suggest that future research should go beyond computation-optimization approaches~\cite{akhtar:20,ali:20,eismann:21a}, given how little computation time contributes to the end-to-end latency of many applications.
The high fraction of external service time shows that fast data exchange between stateless functions remains a key challenge for serverless applications.
Many applications would benefit from low-latency storage solutions such as Pocket~\cite{klimovic:18a}, Shredder~\cite{zhang:19b}, or Locus~\cite{pu:19}.
Finally, efficient function coordination through triggers~\cite{pelle:19,lopez:20} and workflow orchestration~\cite{lin:20b} deserves more research attention given the high transition delays.
\paragraph{Cold start times are best improved by debloating language runtimes.}
Language runtimes should be the primary focus for optimizing cold start latency given their major impact, adding >500\,ms overhead for most applications.
Runtimes were not designed for serverless architectures and recent optimizations for
Java~\cite{blog:Filichkin21}
and 
.NET~\cite{blog:Schaatsbergen21}
achieve large speedups of up to 10$\times$, though sometimes at the cost of more memory usage or larger deployment sizes.
Debloating system stacks~\cite{kuenzer:21} and application dependencies~\cite{soto-valero:20} is another promising optimization motivated by large initialization overheads for applications with large dependency trees (e.g., \app{E} and \app{J}).
Alternatively, serverless developers can select languages with lower runtime initialization overhead, such as Golang~\cite{maissen:20}.

\paragraph{The main cause of tail-latency problems for warm invocations are external services and poorly chosen triggers.}
Latency-critical applications should carefully choose external services and trigger types.
Measurement studies covering different external services can guide the initial selection process~\cite{ustiugov:21a,pelle:19,wang:18,lee:18}.
Our results confirm these findings and identify cloud storage as key contributor to performance variability~\cite{ustiugov:21a}.
Beyond that, sb can provide insights about alternative application implementations.
For example, applications using database services (\app{C}, \app{F}, \app{G}, 
\app{H}) exhibit better tail latency than those using cloud storage (\app{B}, \app{I}, \app{J}).
However, initializing a database connection can add additional cold start delay (c.f., Figure~\ref{fig:latency_warm_cold_diff_p50}).
For asynchronous orchestration, choosing appropriate trigger types is crucial as the cloud storage trigger introduces massive tail latency (e.g., \app{B} in Figure~\ref{fig:latency_warm_p50}).
In contrast, the pub/sub trigger used in \app{C} adds minimal tail latency.
However, queuing time may become an issue as non-HTTP-triggered functions have lower scheduling priority~\cite{tariq:20}.
\begin{table*}
\centering
\footnotesize
\begin{tabular}{@{}llccccccccc@{}}
\toprule
\textbf{Reference}               & \textbf{Focus}                   & \multicolumn{4}{c}{\textbf{Scope}}                  & \multicolumn{2}{c}{\textbf{Invocation Patterns}} & \multicolumn{2}{c}{\textbf{Insights}} \\ \cmidrule{3-6} \cmidrule(lr){7-8} \cmidrule{9-10}
                                 &                                  & apps & func/app & micro & langs & services & concurrent     & trace-based     & white box           & async           \\ \midrule
faas-profiler ~\cite{shahrad:19} & Server-level overheads           & 5    & 1 & 28       & 2         & 0        & \yes          & \no             & \yes                & \no             \\
vHive~\cite{ustiugov:21}         & Cold-start breakdown     & 9    & 1  &    0   & 1         & 1        & \no           & \no             & \yes                & \no             \\
ServerlessBench~\cite{yu:20}     & Diverse test cases         & 4    & 1-7 &   10  & 4        & 1       & \no           & \no             & \no                 & \no             \\
SeBS~\cite{copik:21a}             & Memory size impact  & 10   & 1 &   0     & 2         & 1        & \yes          & \no             & \no                 & \no             \\
FunctionBench~\cite{kim:19}      & Diverse workloads & 8   & 1   &  6    & 1         & 1        & \no           & \no             & \no                 & \no             \\
FaaSDom~\cite{maissen:20}        & Language comparison              & 0    & -  & 5      & 4         & 0        & \yes          & \no             & \no                 & \no             \\
BeFaaS~\cite{grambow:21}         & Application-centric & 1    & 17 &  0    & 1         & 1     & \yes          & \no             & \yes                & \no             \\ \midrule
\textbf{ServiBench (this work)}                   & White-box analysis   & 10    & 1-21  & 0    & 5         & 7        & \yes          & \yes            & \yes                & \yes            \\ \bottomrule
\end{tabular}
\caption{\label{tab:related_work} Summary of related serverless benchmarks.}
\end{table*}

\subsection{Limitations}

Despite careful design, we cannot avoid a small number of limitations in our design and results.
First, \textbf{all results reported in Section~\ref{sec:results} are specific to the AWS serverless platform}. Conceptually, ServiBench enables benchmarking a wider range of cloud providers.
However, cloud providers may not provide detailed tracing information comparable to AWS' X-Ray, preventing in-depth analysis.
Following improved tracing capabilities, we are currently adding support for Microsoft Azure.

Second, \textbf{we do not tune Lambda memory sizes %
for individual benchmark applications or functions}. This is a common decision in benchmark design, to increase fairness of comparison. %
Future research should investigate ideal settings for each application in our benchmark.

Third, \textbf{the load generator currently supports invocation patterns on a per-second granularity}. For simulating extremely bursty workloads, more fine-grained configuration would be necessary, for example to configure a burst that happens within a few milliseconds. This may explain why we observed only a limited impact of different invocation patterns on end-to-end latency in Section~\ref{sec:res_invocation}.

Last, we identify but \textbf{do not address the possible issue of long-term performance changes in cloud settings}. Cloud providers iterate rapidly and also operational policies can change, so performance may change or even vary over time. Future work could address this situation through techniques such as periodic, long-term measurements. %

\section{Related Work}\label{sec:related_work}

This work complements and greatly extends a large body of work on serverless benchmarks and performance analysis.

\textbf{Serverless benchmarks and measurement frameworks}:
Table~\ref{tab:related_work} compares ServiBench with 
the most important serverless benchmarks and performance frameworks.
Our study
\begin{enumerate*} [label=\upshape(\roman*\upshape)]
     
    \item has a wider scope with more applications, functions, and services, and in particular with more than one function or service, 
    
    \item adds realistic applications and invocation patterns based on real-world characteristics~\cite{williams:18,eismann:21-tse,shahrad:20},

    \item does not rely on low-level, server-side tracing, not available for public serverless platforms,
     
    and
     
    \item enables analysis across a variety of situations common in production, including synchronous and asynchronous invocations, end-to-end tracing including external services, and fine-grained white-box analysis.
     
\end{enumerate*}

Closest to our work, BeFaaS~\cite{grambow:21} is an application-centric benchmarking framework, but uses only synchronous function-chains and a single external service (an external database).
BeFaaS enables cloud-agnostic tracing through chained functions, however, the language-specific architecture does not support the analysis of orchestration, queueing, trigger, finalization overheads, and container and runtime initialization.

\textbf{Performance analysis of serverless platforms}:
Performance is an important and commonly studied aspect of serverless computing. Over 100 studies from academia and industry have already appeared~\cite{yussupov:19,hassan:21, scheuner:20-jss, raza:21}. Commonly investigated topics include 
scalability~\cite{kuhlenkamp:20, McGrath:2017}, 
cold starts~\cite{manner18, wang:18}, 
performance variability~\cite{cordingly20,lee:18}, 
instance recycling times~\cite{wang:18, lloyd:2018b}, 
and the impact of parameters such as memory size~\cite{figiela:18, zhang2019}, 
or programming language~\cite{djemame:20, jackson:18}. 
These studies tend to rely on single-purpose micro-benchmarks and rarely utilize tracing data. Further, reproducibility~\cite{taylor:94} remains a big challenge in serverless performance studies, as analyzed recently~\cite{scheuner:20-jss}.
\section{Conclusion}\label{sec:conclusion}

Due to their compositional nature, serverless applications and the platforms executing them are challenging to benchmark.
We designed and implemented ServiBench, an open-source, application-level, serverless benchmarking suite. 
Unlike existing approaches, ServiBench: 
(1) leverages a suite of 10 diverse and realistic applications (importantly, including both synchronous and asynchronous cases), (2) extracts invocation patterns from cloud-provider data and generates realistic workloads,
(3) supports end-to-end experiments, capturing fine-grained application-level performance and enabling reproducible results,
(4) proposes a novel algorithm and heuristics to enable white-box analysis even for asynchronous applications and data produced by (distributed) serverless tracing,
and
(5) supports comprehensive performance analysis for real-world scenarios such as cold starts and tail latency. %

Using ServiBench, we conducted a comprehensive, large-scale empirical investigation of the AWS serverless platform, collecting over 7.5 million execution traces. We observe that median end-to-end latency is most often dominated by external service calls, orchestration, or by waiting for asynchronous triggers. Excessive tail latency is similarly caused more by external services (particularly object storage) than any computation inherent to the serverless applications. Regarding cold starts, our results indicate that investment into simplifying runtime environments or slimming them down (e.g., as Golang does) is the most promising angle to speed up scaling. Finally, our experiments confirm the AWS platform can react effectively to workload differences, even to challenging bursty invocation patterns, for most applications.

In the future, 
ServiBench, and the general serverless benchmarking concepts demonstrated by it, can be used by practitioners to evaluate in-detail performance problems in their own applications or serverless platform of choice. Platform engineers can use our approach and tooling to further improve their offerings. We envision ServiBench to become an integral part of the evaluation of future serverless research contributions, 
through its current features, and as an extensible basis.
\section*{Acknowledgements}
We are grateful to the SPEC Research Group\footurl{https://research.spec.org/} for fruitful discussions and want to thank Johannes Grohmann, Sean Murphy and Jan-Philipp Steghöfer for their contributions.
Special thanks go to our research assistants Simon Trapp and Ranim Khojah for supporting the integration of applications.
We appreciate the generous support of our industry partners enabling large-scale evaluation and detailed investigation of tracing issues with the AWS service teams.
This work was partially supported by the Wallenberg AI, Autonomous Systems and Software Program (WASP) funded by the Knut and Alice Wallenberg Foundation.

\appendix

\section{Replication Package}\label{app:replication}

We provide a detailed replication package with two main goals. First, we want to enable the replication of our results by independent researchers to make the results reproducible. This also allows to track how the reported performance properties evolve over time. Secondly, we want to enable the use of ServiBench in further studies as the reported analysis in Section~\ref{sec:results} covers only a small subset of the analysis enabled by our tool. Further studies could for example analyze different styles of applications (e.g., scientific workflows), different workload patterns (e.g., scheduled jobs), investigate influencing factors for different latency components (e.g., what influences the orchestration delays), or use ServiBench to analyze novel approaches that build on top of public serverless platforms~\cite{eismann:21a, czentye:19, lopez:20}. 

The key component of our replication package is our open-source tool ServiBench, which encompasses the benchmark harness, the trace upscaler for the azure functions traces, and our latency breakdown analysis. It comes with ten realistic serverless applications out of the box and instructions for the integration of additional applications. The ServiBench tool fully automates the application deployment, load generation, metric collection, and trace analysis for the performance analysis of serverless applications. To enable the full replication of the presented results, we additionally include our scripts for the visual inspection of the azure trace dataset, the raw data collected during our measurements, and the scripts to replicate any analysis and figure from Section~\ref{sec:results}.
The replication package is currently available as an anonymous GitHub repository\footurl{https://github.com/ServiBench/ReplicationPackage} and will be archived to Zenodo with a DOI upon publication.

\section{Serverless Application Description}\label{sec:app_description}
This section describes the architecture and functionality of each application introduced in Table~\ref{tab:app_list}. For further details on implementation details, such as usage profile, or service configuration, we refer to our replication package. 

\begin{figure}[h!]
    \centering
    \includegraphics[width=0.25\textwidth]{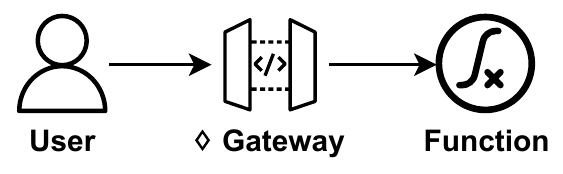}
    \label{fig:app:baseline}
    \label{app:A}
    \caption{\textbf{Minimal Baseline (App A)} \textmd{emulates an HTTP request sent to the API Gateway which triggers a simple serverless function, and returns a  response.
    }}
\end{figure}

\begin{figure}[h!]
    \centering
    \includegraphics[width=0.45\textwidth]{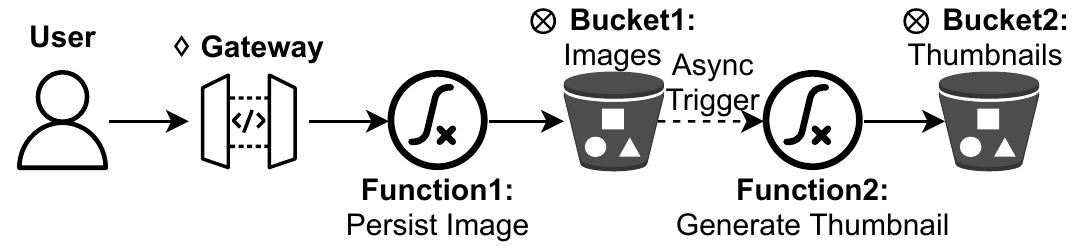}
    \label{fig:app:thumbgen}
    \label{app:B}
    \caption{\textbf{Thumbnail Generator (App B)} \textmd{generates a thumbnail of an image uploaded to a storage bucket. The first function implements an HTTP API to upload an image to a storage bucket. The storage event then triggers a second function to generate a thumbnail of the image and store it in another storage bucket.}}
\end{figure}

\begin{figure}[h!]
    \centering
    \includegraphics[width=0.49\textwidth]{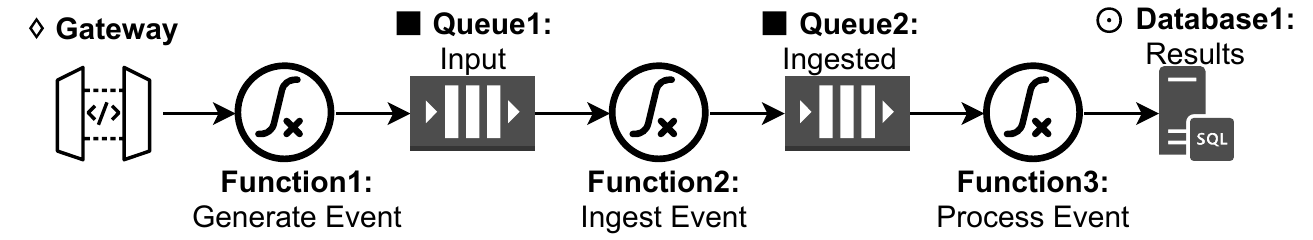}
    \label{fig:app:eventproc}
    \label{app:C}
    \caption{\textbf{Event Processing (App C)} \textmd{generates and inserts events into an input queue. The queue triggers a function which pre-processes the event and places it in the ingested queue. The placement of an event in the ingested queue triggers another function to process the event and store the results in the database.}}
\end{figure}

\begin{figure}[h!]
    \centering
    \includegraphics[width=0.4\textwidth]{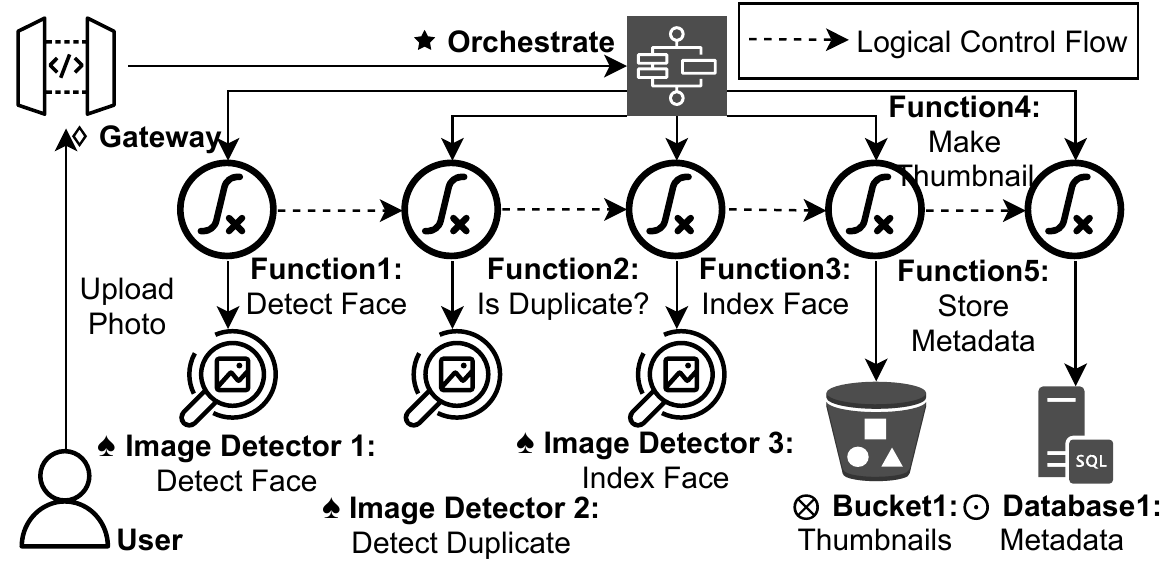}
    \label{fig:app:facerec}
    \label{app:D}
    \caption{\textbf{Facial Recognition (App D)} \textmd{app takes a user uploaded image, extracts a face from it, and detects if the face already exists in the database. If the face does not already exist in the database, the app indexes the face and saves a thumbnail of the face to object storage.}}
\end{figure}

\begin{figure}[h!]
    \centering
    \includegraphics[width=0.35\textwidth]{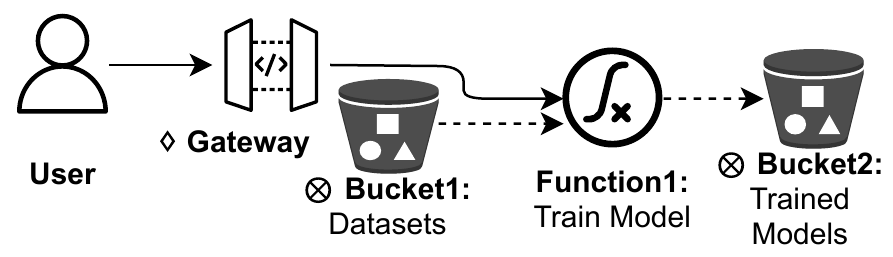}
    \label{fig:app:modeltrain}
    \label{app:E}
    \caption{\textbf{Model Training (App E)} \textmd{application reads training datasets from object storage, trains machine learning models  on those datasets, and stores the trained models in another object storage bucket again. The model is a logistic regression to predict review sentiment scores on the Amazon Fine Food Review dataset.}}
\end{figure}

\begin{figure}[h!]
    \centering
    \includegraphics[width=0.3\textwidth]{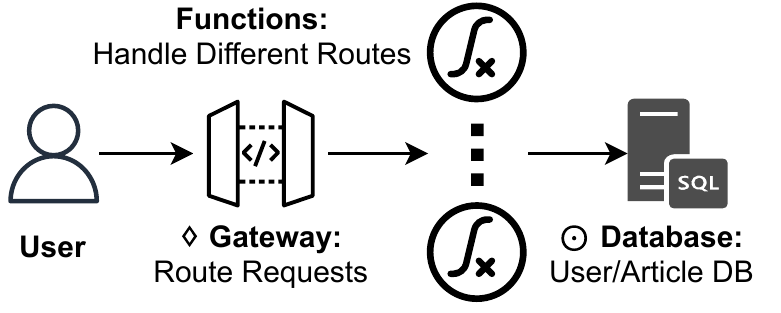}
    \label{fig:app:realworld}
    \label{app:F}
    \caption{\textbf{RealWorld Backend (App F)} \textmd{uses a functions to create, read, update, and delete user and article information stored in a database.}}
\end{figure}

\begin{figure}[h!]
    \centering
    \includegraphics[width=0.4\textwidth]{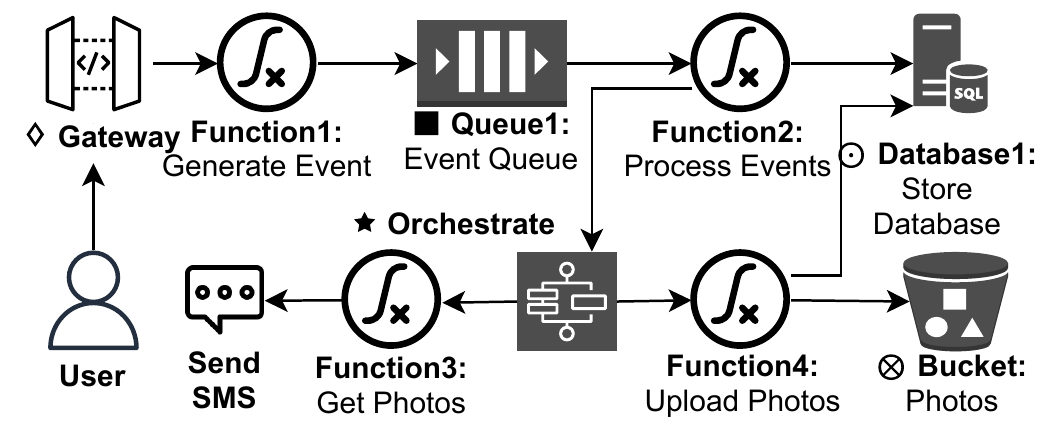}
    \label{fig:app:helloretail}
    \label{app:G}
    \caption{\textbf{Hello Retail! (App G)} \textmd{is a retail inventory catalog application backed by a database. Users can upload product information and categorize products into categories. Supports sending an SMS if a product does not have an image. Uploaded images are stored in object storage.}}
\end{figure}

\begin{figure}[h!]
    \centering
    \includegraphics[width=0.3\textwidth]{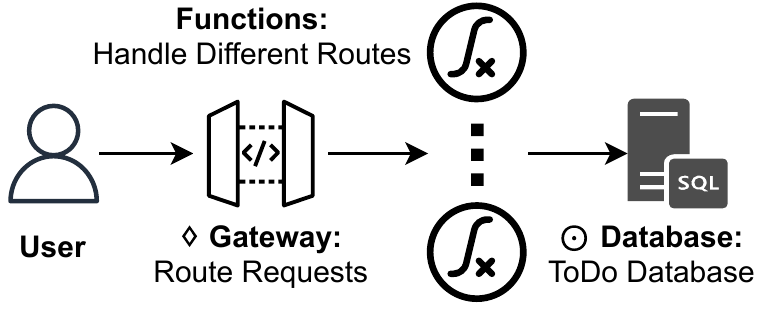}
    \label{fig:app:todo}
    \label{app:H}
    \caption{\textbf{Todo API (App H)} \textmd{is a simple to-do app which uses a FaaS to create, read, update, and delete todos stored in a database.}}
\end{figure}

\begin{figure}[h!]
    \centering
    \includegraphics[width=0.49\textwidth]{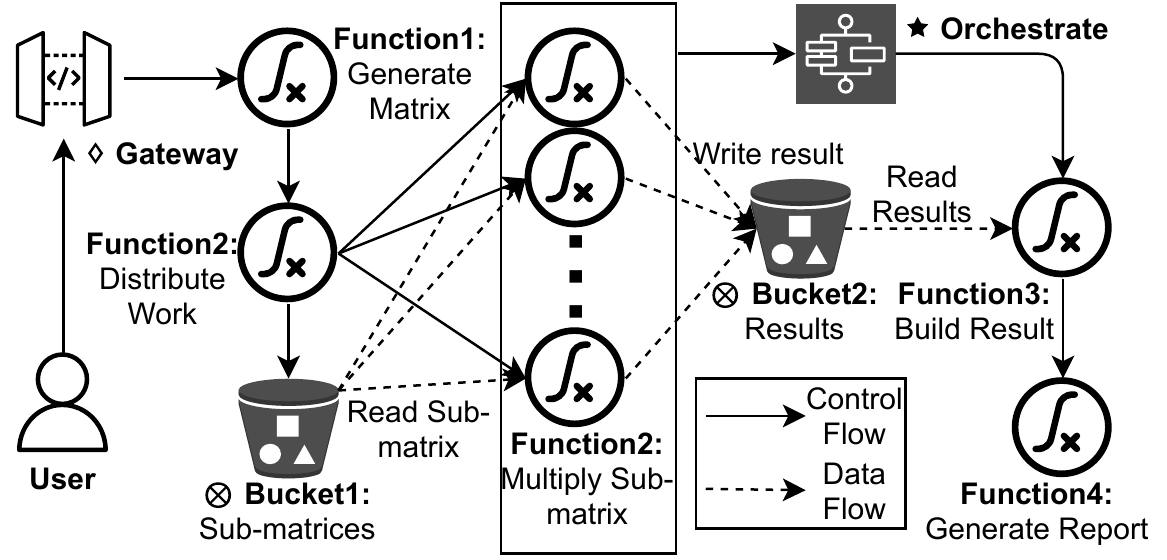}
    \label{fig:app:matmul}
    \label{app:I}
    \caption{\textbf{Matrix Multiplication (App I)} \textmd{generates a random matrix, partitions the matrix, and distributes it for multiplication. Workers perform the multiplication and write the results to S3. The results are then combined to get the final result of the multiplication. The app is directed by an orchestration service.}}
\end{figure}

\begin{figure}[h!]
    \centering
    \includegraphics[width=0.35\textwidth]{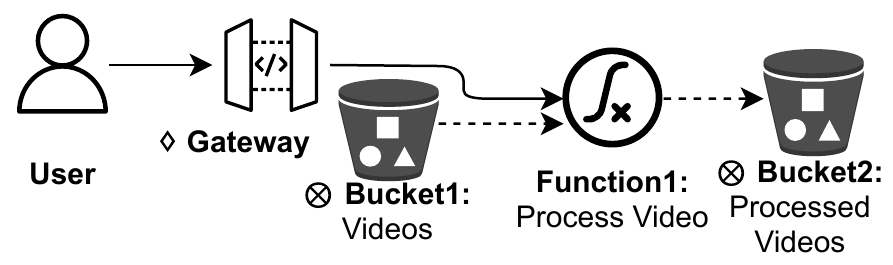}
    \label{fig:app:videoproc}
    \label{app:J}
    \caption{\textbf{Video Processing (App J)} \textmd{application reads videos from object storage, applies a greyscale filter, and transcodes them. The transcoded videos are stored in object storage.}}
\end{figure}

\break
\bibliographystyle{plain}
\bibliography{references}

\end{document}